# A Binary Data Stream Scripting Language

LIHUA WANG AND LUIZ F. CAPRETZ
Department of Electrical and Computer Engineering
University of Western Ontario
London, Ontario, N6A5B9
CANADA
lwang224@uwo.ca, lcapretz@eng.uwo.ca

*Abstract*: Any file is fundamentally a binary data stream. A practical solution was achieved to interpret binary data stream. A new scripting language named Data Format Scripting Language (DFSL) was developed to describe the physical layout of the data in a structural, more intelligible way. On the basis of the solution, a generic software application was implemented; it parses various binary data streams according to their respective DFSL scripts and generates human-readable result and XML document for data sharing. Our solution helps eliminate the error-prone low-level programming, especially in the hardware devices or network protocol development/debugging processes.

## 1  Introduction

The distinction between binary file and text file is only useful to computer users. A binary file is computer-readable but is hard to read for humans such as image files, sound files, compiled computer programs or compressed versions of other files. A text file is human-readable because it contains bytes that can be directly interpreted as characters following one of the standard text schemes (Unicode, ASCII, EBCDIC, etc). However, text files are essentially a special case of binary file and most software systems make no distinction between file types. Only the data format makes the difference.

Data format is the key for determining the physical layout and semantic meanings of the data [1]. Producing a parser to parse an arbitrary data stream according to its data format is a crucial step in data processing.

Currently, a programmer must choose a language, convert the documented data format description into data structures such as C structures; then the data must be read into memory, some operations performed, and the data written back to external storage [2].

The C programming language is the most commonly chosen language for writing such programs, especially to implement real-time algorithms in a system that interfaces with bare hardware devices [3]. Engineers incur time-consuming and error-prone penalties from low efficiency C languages when working at the bit level.

We propose a new scripting language to specify the binary data format from lowest granularity level. This language named Data Format Scripting Language (DFSL) is easy to learn, straightforward to understand, and agile to fulfil a users requirements. The advantages of this solution are:

1. The only tool a user needs is a text editor; no compiler or linker is necessary.
2. No complicated programming is required.
3. The language is easy to use with simple syntax and rich semantic meaning. It is very close to the tabular configuration used to define the data format.
4. It saves developing time because no time-consuming low-level coding is necessary.
5. Many functions are optimised and hidden from the user such as file access and low-level bits manipulation.

We have also developed a generic application to parse the script and the raw data. The user can write a script for a specific data format using DFSL language and the associated application can parse the script and the raw data. The outputs are a meaningful interpretation of the data fields and XML document of the data structure. As a result, this scripting language frees the programmers from tedious tasks of coding input/output routines, and helps eliminate the mishaps of low-level coding.

## 2 Existing Problems

The original motivation came from the need for an easier way to parse and interpret the arbitrary bit streams in hardware device development. Most devices such as drivers and interface cards have control and status registers. The device is controlled by setting and clearing particular bits in the *control register* (see Figure 1), while its status is obtained by examining bits in the *status register*.

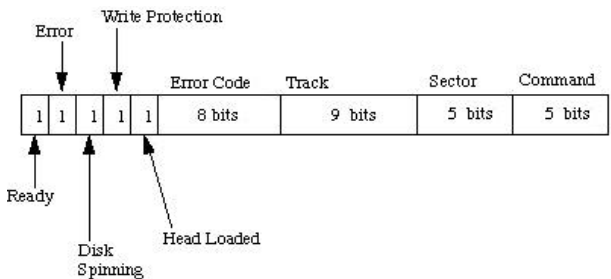

Figure 1. Disk control register

There are many of these registers and each register, or even each bit represents different meanings. It is tedious and error-prone to check the value manually. Sometimes, the engineers develop special programs to parse the data. However, it is also error-prone to process low-level bits using C and it is not worth the time to implement the program since the user requirement changes frequently.

The network application developers face the same problem when writing network data processing code. Interpreting network packet is the key to most important network applications including web server testing, network traffic monitoring, network firewall checking, and so forth.

The packets are usually in standardized formats such as TCP/IP, FTP and SSL. Due to the complexity of networking protocol, and the bit-oriented feature of the networking data stream, it is complicated to write a program using C languages to interpret the data stream.

The main problems are byte-alignment constraints, byte-order difference, field dependency & dynamic typing, and system-dependent word-size.

### 2.1 Byte-Alignment Constraints

The data type system of C language is byte-based, meaning it applies byte-alignment restriction. This assumes the storage media and the input/output routines have a minimum unit of one byte [4]. Programmers are conditioned to think of memory as a simple array of bytes.

Data types *char, int* and *double* are primitive types and as well as the other data types of C language, they are all integer multiples of one byte. All the operations are based on these types. However, the real machine-level atomic unit is the bit. When the required granularity is less than a byte (1 byte = 8 bits) and we need bit-fields with arbitrary length, the byte-alignment can become a barrier for fast programming.

Generally, one of three techniques is used to manipulate bit fields:

- Pointer arithmetic and bit masking/shifting: (the << and >> bitwise shift operators).
- Macro defines: some single or multiple bit macros that shift or mask the appropriate bits to get the desired result.
- Bit-field packing structure: the structure construct with bit fields specified by a post-declarator colon and integer field width.

We can use pointer arithmetic to access the memory and bit masking and shifting to achieve the bit operations. However, the pointer arithmetic and bit operations of C language are notorious for their tedious and error-prone nature. Moreover, the semantics end up buried in parsing code and the difficulty of reading the code can make the maintenance even harder.

The macro define is just a set of commands of pointer arithmetic and bit shifting and masking. C language also provides a practical feature called bit fields which automatically packs the bit fields as compactly as possible and provides that the maximum length of the field is less than or equal to the integer word length of the computer, However, bit fields lack portability between platforms because some bit field members are stored left to right while others are stored right to left. Thus the existing approaches are not efficient enough to deal with low-level coding either because of the tediousness and error-prone nature or because of the byte-order problem.

### 2.2 Byte-Order Difference

The word "endianess" describes the method used to represent multi-byte integers in a computer system. There are two types of endianess: big-endian refers to the method of storing the most significant byte of an integer at the highest byte address, and little-endian is the opposite. Consider a number 0x1234 declared as a *short int,* which consists of two bytes. Its first byte stored in the big-endian system is 0x12, but 0x34 in the little-endian system.

Different endianess methods may apply for different machine's architecture, and may cause problems when application runs across different systems. The reason why the bit-field packing structure of C language is lack of portability between platforms is also caused by the endianness difference.

## 2.3 Field Dependency and Dynamic Typing

A data type is a name or label for a set of values and some operations can be performed on that set of values [4]. Some formats and especially some protocol headers can contain fields whose values or sizes depend on the value of a previous field. For instance, the Options field in the IP header can occupy between 0 to 40 bytes depending on the value of the previous field IHL (Internet Header Length) [3]. These fields cannot be defined using static types in C *struct*. Since static types cannot represent variable-sized fields. We need to use dynamic typing to solve this problem.

## 2.4 System-Dependent Word Size

Another problem concerns the system dependent computer word size. The length of an integer (type 'int') traditionally depends on the length of the computer word. For instance, it is 16 bits long in MSDOS, whereas in 32-bit systems (like Windows 9x/2000/NT) it is 32 bits long (4 bytes). The ambiguous type size may produce different results.

Before proceeding with the idea of developing a new language, some people may ask why other existing technologies are not used. The question arises as to whether existing libraries can be used. Firstly it costs time and labour to learn the library interface but it may lack the necessary features we need since the former developers did not foresee a need for different features. On the other hand, it may not be possible to find the library needed for those non-standard or user-defined data format. This frequently occurs in new data format development, and especially when developing and testing new networking protocols.

The problems brought up thus far are all related to the C language. Why other high-level languages are not used (especially those support polymorphic type, such as Haskell, Standard ML)? The reason is that it takes time to learn a new programming language and the user may just need to use a fraction of its functions and a running environment.

Another frequently asked question is why not using XML alone. XML (eXtensible Markup Language) is a meta-language that is a way to define tag sets [5]. To access an XML document file from a program, you can either parse the tag structure in your own code or using one of two standard APIs to invoke parsers to do it for you. The two APIs are DOM (Document Object Model) and SAX (Simple API for XML) [6]. These APIs are still based on byte type system and so far they did not focus on bit field's specification.

# 3 Method Prototype

The architecture of our system is depicted in Figure 2. The *Specified Data Format* on the most left can be any user-defined data format or standardized data format, which is usually described in a human-understandable language.

The *Data Format Scripting Language (DFSL)* has been developed. This language can be used to translate the documented data format into *Data Format Script*, which is the interface between the users and the computer.

Then we have the *Data Format Script* and the *Binary raw data*. They are the two actual inputs for our generic *Data Parsing Application,* which is the main component of our project.

The parsing application can parse the *Data Format Script* and execute any command in the script. The output is *Data Interpretation,* which includes the meaning of each bit field and its value and a XML documentation of the data structure.

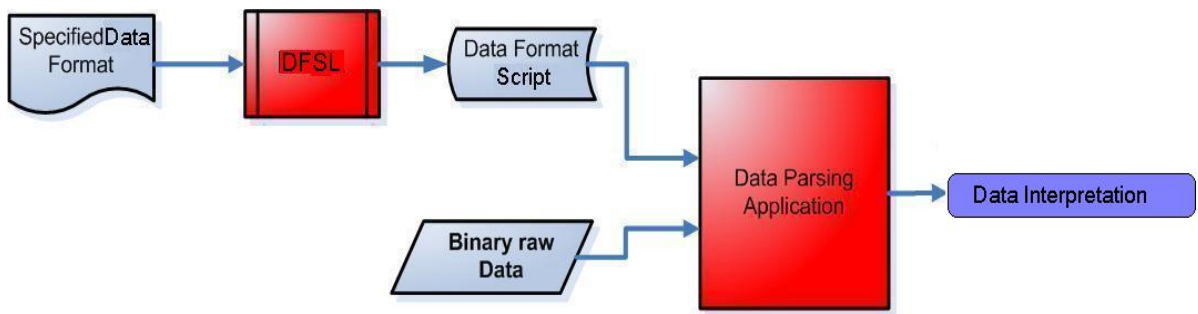

Figure 2. Data interpreting method prototype

This method is not limited to any particular data format. Once the data format is established, all the data with that format can be parsed.

# 4 Data Format Scripting Language

Data Format Scripting Language has been developed as a scripting language to interpret a binary data stream. DFSL has special features that can simplify a user's specification and some simple commands that can facilitate bit field operations. We introduce these items in this section by examples.

## 4.1 Lexical Elements

Lexical elements are basic building blocks of programming languages. The basic elements are:

- Number token: a number token can be an unsigned integer in decimal notation, e.g., 2, 77, 2356 or in hexadecimal notation, e.g., 0x3, 0x b3f5; or a real number, e.g., 123.45, 0.023e-5.
- Word token: the word token is case insensitive. A word token can be an identifier, e.g., aaa, number; or a reserved word, e.g., if, switch.
- Domain token: a domain token starts with a dollar sign ($) followed by letters, underscores or digits, e.g., $SIP_Packet.
- Sub-Domain token: a sub-domain token starts with a percent sign (%) followed by letters, underscores or digits, e.g., %height.
- String token: a string token consists of a sequence of characters enclosed by a pair of quotation marks, e.g., "A string literal".
- Comment: they are started with double slash "//" and goes to the end of the line. Comments and white-space characters are bypassed during the scanning process.
- Operators : most special tokens are used as operators, such as "+", "*", ">=".
- Expressions: expressions are composed of one or more constants, variables or function calls and zero or more operators.
- Selection statements: if and switch statements.
- Loop statements: while, do-while, and for statements.

## 4.2 Layered Architecture Specification

When describing the layout of the data format, people usually use a tabular form, which represents a layered architecture (or called as hierarchical architecture). In our language, we use grouping and sequencing to specify the layered structure of the data format. Grouping can gather all the component members in the structure body, and the sequence of individual fields indicates the physical layout of the bit stream.

The basic concepts of layered architecture can be summarized as following:

- The data format is considered as a layered structure.
- All the data with the same data format are in one general domain and can be represented by one domain name.
- Top-down point of view: the top-most domain is called the root domain that has a set of lower-level domains.
- The lower-level domains are known as sub-domains and have their own structure.
- A sub-domain can be further divided into even lower-level sub-domains or elementary components.
- The elementary component in a domain or a sub-domain can be one bit or a fixed size array of bits.

A representative example is parsing an ICMP (Internet Control Message Protocol) packet. The hexadecimal dump in Figure 3 is an ICMP ECHO response message (packet) [7].

```
 0: 0800 2086 354b 00e0 f726 3fe9 0800 4500
16: 0054 aafb 4000 fc01 fa30 8b85 e902 8b85
32: d96e 0000 45da 1e60 0000 335e 3ab8 0000
48: 42ac 0809 0a0b 0c0d 0e0f 1011 1213 1415
64: 1617 1819 1a1b 1c1d 1e1f 2021 2223 2425
80: 2627 2829 2a2b 2c2d 2e2f 3031 3233 3435
96: 3637
```

Figure 3. Hex dump of ICMP ECHO response packet

The packet can be broken into the following protocol elements: Ethernet header, IP header and ICMP datagram as shown in Table 1:

Table 1. ICMP packet structure

| Ethernet Header |
| --- |
| IP Header |
| ICMP datagram |

The first part is the Ethernet header, which includes three fields (see Table 2):

Table 2. Ethernet header structure

| MAC Destination Address | ( 0 - 5): six bytes |
| --- | --- |
| MAC Source Address | ( 6 - 11): six bytes |
| Ethernet Type Field | (12 - 13): two bytes |

The second part is the IP header whose structure is shown in Figure 4. We can see some of the fields are not integer multiple of one byte: Version is 4 bits; IHL (Internet Header Length) is 4 bits; Flags (Various Control Flags) is 3 bits; FragmentOffset is 13 bits and the detailed definitions are in *RFC 791* [8]. The last part is the ICMP datagram, which is the ICMP header followed by the payload data.

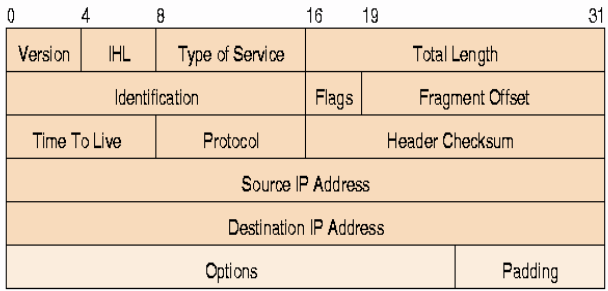

Figure 4. IP header structure

DFSL language can be used to parse such layered data structures without resorting to complicated programming. The user only needs to write a simple script to describe the layout of the packet and leaves the data parsing task to the DFSL application.

As shown in Figure 5, the domain structure definition has the form *$domain_variable := { ... }*. All the components of an ICMP response packet structure are defined and grouped using a pair of matched braces under the domain variable *$ICMP_response*. The sequence of the components indicates the position of those fields in the real data stream.

Sub-domain variables (starts with %) are used to store the value of the elementary component or the link to other domain. An assignment statement with the form *%sub-domain = right_value* is used to define the sub-domain variable. The *right_value* can be a standard command or another domain structure definition. As is seen from the sample code, most variables in the IP header domain (which is labelled with *$IP_header*) get the bits or byte value (e.g. %version = getBit 4) and they were called elementary component; else wise, some components are assigned to other domain structure (e.g., %source = $ipAddress).

The layered architecture means that we can use grouping to specify the data structure and sequencing to describe the layout of the data. The sequence of items determines the execution order. When the parser finds a sub-layer label, it traverses the sub-layer and resolves the fieldnames before returning, subsequently it executes the next entry.

```
█ ▼ "H:\2005-Project\!October\Book2-1\Deb

¶Page 1    ICMP.txt    Sat Oct 29 20:46:07 2005

 1: $ICMP_response = getFile ( "icmp.dat");
 2: $ICMP_response := {
 3:     %Ether_header   = $Ether_header;
 4:     %IP_header      = $IP_header;
 5:     %ICMP_header    = $ICMP_header;
 6: }
 7: $Ether_header := {
 8:     %destination    = $MAC_address;
 9:     %source         = $MAC_address;
10:     %type           = getByte 2;
11: }
12: $MAC_address := {
13:     %vendor         = getByte 3;
14:     %serialNumber   = getByte 3;
15: }
16: $IP_header := {
17:     %version        = getBit 4;
18:     %IHL            = getBit 4;
19:     %TOS            = getByte;
20:     %identification = getByte 2;
21:     %flag0          = getBit;
22:     %flag1          = getBit;
23:     %flag2          = getBit;
24:     %offset         = getBit 13;
25:     %time2live      = getByte;
26:     %protocol       = getByte;
27:     %checksum       = getByte 2;
28:     %source         = $ipAddress;
29:     %destination    = $ipAddress;
30: }
31: $ipAddress := {
32:     %first  = getByte;
33:     %second = getByte;
34:     %third  = getByte;
35:     %forth  = getByte;
36: }
37: $ICMP_header := {
38:     %Type     = getByte;
39:     %code     = getByte;
40:     %checksum = getByte 2;
41: }
42:
Press any key to continue_
```

Figure 5. Script for ICMP packet

## 4.3  Bits Operation Solution

One of the advantages of using our language is that it hides the tedious bit operations from the user. In C programming language, the bit field is limited to the boundaries of the underlying object that is of the fundamental C type. That means the bit field may not be wider than the underlying object and no bit field should overlap the underlying variable boundaries. Thus the C structure cannot be used to extract some boundary-crossing bit field in an arbitrary binary record. More complicated operations to extract both ends of such a bit field and put them back together are necessary [9].

DFSL language regards the data stream as an unsigned character array of arbitrary length. Its routines extract a bit field of specified arbitrary length at a certain location, independent of the char boundaries along the array. So we avoided byte-alignment and byte-order problem when read the bits. Its straightforward use achieves the desired bit fields by using command *getBit, getByte, seeBit and seeByte*. Those commands return integer value of the bit field.

The syntax of this series command is as follows:

- **getBit** *count*

Semantic: Read the number of *count* bits from the current position in the bit stream. If offset is not given, default read one bit. The *count* can be a numeric number or an existing variable's value, e.g., getBit 4.

- **getBit** *@position , count*

Semantic: Read the number of *count* bits from the specified position that is indicated by *@position*. If offset is not given, default read one bit. The *count* can be a numeric number or an existing variable's value, e.g., getBit @15, 3.

- **getBit** *start ~ stop*

Semantic: Read bits from *start* position to *stop* position, e.g., getBit 15 ~ 9.

- **getByte** *count*

Semantic: Read the number of *count* bytes from the current position. The *count* can be a numeric number or an existing variable's value, e.g., getByte 3.

- **seeBit**

Semantic: **seeBit** shares the same syntax as **getBit**. It can only preview the bits without moving the bit pointer from the current position.

- **seeByte**

Semantic: **seeByte** shares the same syntax as **getByte**. It can only preview the bytes without moving the bit pointer from the current position.

A pointer was assigned to keep the current bit position in the binary data stream. Originally it will point to the first bit of the stream. After the parsing process begins, the program will extract certain number of bits from the bit sequence whenever it encounters one of these commands. And the pointer moves the next position depending on the command. We bring in *seeBit* function that shares the same function as *getBit* but protects the continuity of the original data stream.

## 4.4 Constraint and Control

As long as the bit stream is not exhausted, the data stream can be chopped and assigned to a field variable. However, there are always constraints on the data, and the extracted data field may also be operated.

We use a keyword *where* to follow the current layer structure definition, and the statements braced in the *where* clause can do the real job, either be applying constraints on bit field or performing operations on the extracted data fields while having no effect on the data layout.

An example for parsing a PMD (Performance Motion Device) device register follows. Eight serial bit-fields appear in a 16-bits binary record and each bit-field occupies several digit places (see Table 3).

Table 3. PMD structure

| PMD Configuration 3 | | | |
|---|---|---|---|
| 15 ~ 11 | 10 ~ 9 | 8 | 7 |
| Tx Power Cutback Value | Tx Power Cutback Mode | SBM Disable | Single Upstream Disable |
| 6 | 5 | 4 | 3~0 |
| China loop | OL Disable | ROL Disable | Hybrid Select |

The binary digits change in real time under different circumstances. DFSL language is used to specify the arbitrary binary records in a C type-independent and type boundary-independent way. The eight variables grouped under the $PMD3 domain get the bit sequence from the data stream according to their sequence. A sample code is shown in Code 1.

```
$PMD3 = 0x9351 ;
$PMD3 := {
    %TxPowerValue = getBit 15 ~ 11 ;
    %TxPowerMode  = getBit 10 ~ 9 ;
    %SBM          = getBit @8 , 1 ;
    %SUpstream    = getBit @7 , 1 ;
    %ChinaLoop    = getBit @6 , 1 ;
    %OLDisable    = getBit @5 , 1 ;
    %ROLDisable   = getBit @4 , 1 ;
    %HybridSelect = getBit @3 , 4 ;
} where {
  println ("Tx Power Cutback Value =",%TxPowerValue);
  print ("Tx Power Cutback Mode  = ", %TxPowerMode);
  switch (%TxPowerMode){
      case 0: println(" -- No Tx Power"); break;
      case 1: println(" -- Manual Tx Power Cutback");
          break;
      case 2: println(" -- Automatic Tx Power Cutback");
          break;
      default: println(" -- Reserved");
  };
  print ("SBM Disable            = ", %SBM);
  if (%SBM == 0)
      { println(" -- Enable SingleBitMap"); }
  else
      { println(" -- Disable SingleBitMap"); };
  ... ...
  print ("HybridSelect           = ", %HybridSelect);
  switch(%HybridSelect) {
      case 0: println(" -- Default"); break;
      case 1: println(" -- GPIO in tri-state mode");
          break;
      default: println(" -- Reserved");
  };
}
```

Code 1. Script segment for PMD register

Figure 6. Output of PMD script

The output for this script is illustrated in Figure 6. The data is parsed according to the data format and interpreted based on the extracted data value.

## 4.5 Casual Interpretation

We use the concept of domain and sub-domain to construct the layered architecture for complicated data format. However, for some straightforward formats, it is more practical to interpret the result right after getting the bit field's value. So we allow the output routine be placed within the structure definition. Users are free to add operational code once they get the values they need. This is suitable for simple and straightforward format, and will not cause much trouble for future reading.

Thus far we have briefly exposed the main feature of the DFSL language. As a scripting language still in development, the DFSL project is an open-ended project where we try to describe more and more data formats. The associated application is a generic software needed to parse this language. Using the application and the script, users can easily interpret a bit stream.

## 5 Implementation

We have implemented an interpreter for DFSL. The DFSL interpreter works in four steps:

1. It takes in the input domain definitions and produces a parse tree consisting of a node for each item.
2. It performs semantic processing to resolve domain variable and propagate the parse tree, so that structures of fixed size can be recognized.

3. It produces an elaborated intermediate presentation for each node in the definition.
4. It interprets the elaborated intermediate representation and generates output.

The first step is carried out by the scanner. It reads in the textual script written in DFSL breaking it into tokens. A basic parse tree consisting of a node for each item (domain and sub-domain) is constructed and each node contains important attribute information of the item.

The second and third steps are both part of the syntax semantic analysis which is controlled by the parser. The parser performs bottom-up "node propagation" using depth-first traversal, so that the domains of fixed size are annotated with their size.

A domain has a fixed size if:

- It is a bit,
- It is a fixed size array of bits,
- It is a binary string literal,
- It is a definition list (:=) consisting only of fixed size member.

The parser resolves the sub-domain variable if it is a fix-sized field or is linked to a domain which only consists fix-size member. If the sub-domain node meets these requirements, the children-leaves of the linked domain will become the children of this sub-domain. Meanwhile the elaborated intermediate presentation for each node is produced.

In the last step, the executor traverses all the terminal leaves in depth first order. It extracts the binary digits from the data stream and interprets the intermediate presentation on each node. The XML document of the data structure is generated as well as the interpreted output.

At the present time, the DFSL language focuses on parsing the binary data stream and providing facilities to manipulate data fields and user-defined variables. The aim is to give users the flexibility to put the integral-constraints on the data field. It has been tested with several data formats but further extension might be required to prove its effectiveness for extremely complicated data formats.

A number of experiments were conducted on various data format to test and validate the proposed approach and the tool. Some were popular or historical data formats and some were just used in special field. The experiments showed some limitations of the system, but overall the system proved to be applicable to solve a wide range of problems dealing with data stream.

# 6 Conclusions

After investigating problems in a real-world industry environment, a practical solution has been achieved to interpret raw data: regarding any kind of data as bit-stream, using a scripting language to describe the data format, and applying a software application to interpret the raw data according to the script.

This solution is relatively new and no similar application presently exists. The main contributions of our work are:

- The approach solves some existing problems when C language is used, such as byte-alignment constraints, byte-order difference, field dependency and dynamic typing difficulty, and system-dependent word-size problem.
- The approach can be applied widely, anywhere involving raw data, including hardware device development/debugging, network protocol processing or binary file analysis.
- The DFSL language is readable non-specialist and easy to learn.
- The system reduces the complexity of low-level program.
- XML document can be generated for data sharing between applications.
- The most valuable usage of this language is in interpreting data format that is not standardized or at least not defined in public, such as developing or testing new hardware.

The DFSL system can assist the interpretation of machine-level binary digits without complicated programming. As a result, it frees programmers from the tedious task of coding input/output routines, and eliminates the distractions from architecture-dependent problems thus making it possible to achieve higher development productivity.